\documentclass[12pt]{article}

\usepackage{graphicx}
\usepackage{geometry}
\renewcommand{\baselinestretch}{1.5}

\usepackage{amssymb}
\usepackage{amsmath,bm}

\newcommand{\bey}{\begin{eqnarray}}
\newcommand{\eey}{\end{eqnarray}}
\newcommand{\beq}{\begin{equation}}
\newcommand{\eeq}{\end{equation}}

\newcommand{\boldm}[1] {\mathversion{bold}#1\mathversion{normal}}

\newtheorem{thm}{Theorem}[section]

\begin{document}

\vspace*{0in}

\begin{center}

{\large \bf A constrained minimum criterion for model selection}

\bigskip
Min Tsao%\footnote{CONTACT Min Tsao, mtsao@uvic.ca.}
\\{\small Department of Mathematics \& Statistics, University of Victoria, Canada}

\end{center}

\bigskip

{
\noindent {Abstract:} We propose a hypothesis test based model selection criterion for the best subset selection of sparse linear models. We show it is consistent in that the probability of its choosing the true model approaches one and the parameter values of its chosen model converge in probability to that of the true model as the sample size goes to infinity. This criterion is capable of controlling the balance between the false active rate and false inactive rate of the selected model, and it can be applied with other methods of model selection such as the lasso. We also demonstrate its accuracy and advantages with a numerical comparison and an application.
\bigskip

\noindent {\bf Keywords:} {Sparse linear models; Model selection criterion; Hypothesis test; Best subset selection; Model selection consistency.}
}

%%%%%%%%%%%%%%%%%%%%%%%%%%%%%%%%%1111111111111111111111111111111111111111
\section{Introduction}

Consider the linear regression model
\begin{equation}
\mathbf{y}=\mathbf{X}\bm{\beta} +\bm{\varepsilon}, \label{m0}
\end{equation}
where $\mathbf{y}=(y_1,\dots,y_n)^T$ is the response vector, $\mathbf{X}=[\mathbf{1}, \mathbf{x}_1,\dots,\mathbf{x}_{p}]$ is the design matrix, $\boldsymbol{\beta}=(\beta_0, \beta_1,\dots,\beta_{p})^T$ is the unknown vector of model parameters and $\boldsymbol{\varepsilon}\sim N(\mathbf{0}, \sigma^2\mathbf{I})$ is the vector of random error. A variable $\mathbf{x}_i$ is said to be active if its parameter $\beta_i\neq 0$ and inactive if $\beta_i= 0$. Suppose model (\ref{m0}) is sparse in that some of variables are inactive. 
The best subset selection is often used to help identify the true model that contains only and all active variables.  

There are a number of model selection criteria for the best subset selection in both frequentist and Bayesian flavours and with a wide range of motivations.  The most commonly used ones include the adjusted $R^2$,  Akaike information criterion \textsc{aic} (Akaike, 1974), Bayesian information criterion \textsc{bic} (Schwarz, 1978), Mallows' $C_p$ statistic (Mallows, 1973) and cross-validation. The \textsc{aic} and \textsc{bic} are penalized log-likelihood measures of a model with a penalty term proportional to the model size. The adjusted $R^2$ is a model size adjusted goodness-of-fit measure. Mallows' $C_p$ statistic and cross-validation are based on measures of prediction error of a model. Under a given criterion, usually the model that minimizes or maximizes the corresponding measure is selected. The sparsity in the selected model is not explicitly pursued and from this standpoint it is only a by-product of the model selection process. There is a large body of literature on model selection criteria and methods. For a comprehensive review, see Kadane and  Lazar (2004), Ding, Tarokh and Yang (2018) and Miller (1990).

In this paper, we propose a new model selection criterion based on a hypothesis test which pursues sparsity in the selected model explicitly.
In frequentist inference, it is well known that a hypothesis test can be used to derive a point estimator and a confidence interval. The Hodges-Lehmann estimators (Hodges and Lehmann, 1983) derived from various tests are highly efficient. For estimating the unknown true parameter vector of model (\ref{m0}), which we denote by $\bm{\beta}^t$, the least squares estimator may also be viewed as the Hodges-Lehmann estimator derived from the likelihood ratio test for $H_0: {\bm{\beta}^t=\bm{\beta}}$ as it is of the highest $p$-value and in the centre of this test induced confidence region for $\bm{\beta}^t$. However, this estimator cannot be used for model selection as it does not generate sparse solutions. In order to find a sparse solution, we derive from the likelihood ratio test a new sparse estimator for $\bm{\beta}^t$ by considering the collection of models with $p$-values above a given level. This collection is also the set of models not rejected by the test at the given level. Our proposed model selection criterion is to choose from this collection the most sparse member as the estimator for $\bm{\beta}^t$. Let ${\cal M}=\{{M}_j\}^{2^p}_{j=1}$ be the collection of $2^p$ subsets of the $p$ variables in the full model (\ref{m0}) where each $M_j$ represents a subset or model. Let $\hat{\bm{\beta}}_j$ be the least squares estimator for model ${M}_j$. 
Using the pseudo-norm $\|\bm{\beta}\|_0$ that counts the number of non-zero elements in a vector $\bm{\beta}$, the proposed criterion may be formulated as the solution to the following constrained optimization problem, 
\beq
\underset{\bm{{\cal M}}}{\text{minimize}}       \|\hat{\bm{\beta}}_j\|_0
\mbox{\hspace{0.1in} subject to \hspace{0.01in} }  \lambda(\hat{\bm{\beta}}_j) \leq \kappa_n,  \label{form1}
\eeq
where $\lambda({\bm{\beta}})$ is the likelihood ratio test statistic defined in (\ref{lrs}) and $\kappa_n$ is the critical value that depends on the significance level $\alpha_n$ of the test. Because of  (\ref{form1}), we call the criterion the constrained minimum criterion or \textsc{cmc} for short. We call the solution vector to the optimization problem (\ref{form1}) the \textsc{cmc} solution and its corresponding model the \textsc{cmc} selection. There may be multiple solution vectors, and when this happens we choose the one with the highest likelihood as the \textsc{cmc} solution.

Unlike the existing criteria, the \textsc{cmc} explicitly seeks maximum sparsity from plausible models, i.e., models not rejected by the underlying test at level $\alpha_n$. As such, it uses the sparsity condition more aggressively and directly than the existing criteria. We show that with a properly chosen $\alpha_n$ the \textsc{cmc} solution converges in probability to the true parameter vector $\bm{\beta}^t$ and the probability that the \textsc{cmc} selection is the true model approaches one when the sample size $n$ goes to infinity. We compare the \textsc{cmc} with three popular model selection criteria through a simulation study to demonstrate its excellent accuracy and its ability to control the balance between the false active rate and false inactive rate through the choice of $\alpha_n$. Furthermore, we demonstrate its consistency and observe that with a properly chosen $\alpha_n$ the  \textsc{cmc} selection converges to the true model faster than the \textsc{bic} selection. We also apply the \textsc{cmc} to perform model selection and estimation for a prostate cancer data set. The \textsc{cmc} is defined for the best subset selection as the optimization in (\ref{form1}) is over the set of all models ${\cal M}$, but it can be easily extended for other methods of model selection such as the lasso by replacing the set of models  ${\cal M}$  in (\ref{form1}) with the sequence of models found along the lasso path. The underlying idea of inverting a hypothesis test for model selection by choosing the most sparse model not rejected by the test generalizes the Hodges-Lehmann estimator in the context of model selection, and it can also be applied to model selection problems beyond linear models. 

The rest of this paper is organized as follows. In Section 2, we present theoretical results about the  \textsc{cmc} and the numerical comparison. We also give recommendations on choosing the $\alpha_n$ level, including a default level.  In Section 3, we apply the \textsc{cmc} to analyse the prostate cancer data. We conclude with a few remarks in Section 4. The proof of the theoretical results is relegated to the Appendix.

%2222222222222222222222222222222222222222222222222222222222222222222222222222222222222222222222222222222

\section{The constrained minimum criterion}

We assume that $n>p$ and variables $\mathbf{x}_1,\dots,\mathbf{x}_{p}$ are linearly independent so that the least squares estimator for the unknown $\boldsymbol{\beta}^t$ of the full model (\ref{m0}),
\begin{equation}
\hat{\boldsymbol{\beta}}= (\mathbf{X}^T\mathbf{X})^{-1}\mathbf{X}^T\mathbf{y}, \label{ole}
\end{equation}
is available. Although a model $M_j$ may have fewer than $p$ variables, we view its least squares estimator $\hat{\bm{\beta}}_{j}$ as a $q=p+1$ dimensional vector having zeros in elements 
corresponding to variables not in ${M}_j$, e.g., suppose ${M}_2$ contains only $\mathbf{x}_1$ and $\mathbf{x}_2$, then elements of $\hat{\bm{\beta}}_{2}$ representing the coefficients of $\mathbf{x}_3,\dots,\mathbf{x}_{p}$ are all set to zero. Similarly, we view the true parameter vector $\bm{\beta}^t$ as a $q$ dimensional vector. 
%The null hypothesis for the likelihood ratio test underlying the \textsc{cmc} formulation (\ref{form1}) can now be written as $H_0:$ $\bm{\beta}^t=\hat{\bm{\beta}}_j$.

We now present the asymptotic properties of the \textsc{cmc}. In the following, for a constant $\alpha \in (0,1)$ we use $F_{1-\alpha, q, n-q}$ to denote the $(1-\alpha)$th quantile of the $F$ distribution with $q$ and $(n-q)$ degrees of freedom. We need the following regularity conditions on the design matrix $\mathbf{X}$:
\beq
D_n=\frac{1}{n} \sum^n_{i=1}  \mathbf{x}_{ri}\mathbf{x}_{ri}^T \rightarrow D, \label{cond1}
\eeq
where $\mathbf{x}_{ri}$ is the $i$th row of $\mathbf{X}$ and $D$ is a $q\times q$ positive definite matrix, and
\beq
\frac{1}{n} \max_{1\leq i \leq n} \mathbf{x}_{ri}^T\mathbf{x}_{ri} \rightarrow 0.  \label{cond2}
\eeq
Conditions (\ref{cond1}) and (\ref{cond2}) are commonly used to ensure the least squares estimator of the full model $\hat{\bm{\beta}}$ is consistent and asymptotically normal; see, for example, Knight \& Fu (2000). Under these conditions, we have
\beq
\sqrt{n}(\hat{\bm{\beta}} - \bm{\beta}^t) {\longrightarrow} N(\mathbf{0}, \sigma^2 D^{-1}). \label{asynorm}
\eeq
Denote by $M^*_j$ the unknown true model containing only and all active variables and  by $\hat{\bm{\beta}}_{j}^*$ the least squares estimator for $M^*_j$. The following theorem gives the consistency of the \textsc{cmc}.

% Theorem Theorem Theorem ---------------------------------------------------------------------------------------------------------------
\begin{thm} \label{thm1}

Suppose conditions (\ref{cond1}) and (\ref{cond2}) hold. Let $\alpha_n$ be a sequence in $(0,1)$ such that $\alpha_n \rightarrow 0$ and $F_{1-\alpha_n,q, n-q}=o(n)$, and let
\beq 
\kappa_n=q F_{1-\alpha_n,q, n-q}. \label{kappa} 
\eeq 
Denote by $\hat{\bm{\beta}}_{\alpha_n}$ be the \textsc{cmc} solution of (\ref{form1}) with $\kappa_n$ defined in (\ref{kappa}). Then, 
\beq \hat{\bm{\beta}}_{\alpha_n} \stackrel{p}{\longrightarrow} \bm{\beta}^t \mbox{\hspace{0.3in} } \label{convg1} \eeq
as $n\rightarrow +\infty$, and the probability that the true model is chosen satisfies
\beq \lim_{n\rightarrow +\infty}P(\hat{\bm{\beta}}_{\alpha_n} = \hat{\bm{\beta}}_{j}^*)=1 \label{convg2}. \eeq

\end{thm}

\vspace{0.2in}
A proof of this theorem and a remark about the condition on $\alpha_n$ are given in the Appendix.  
Using the \textsc{cmc} for model selection amounts to using the  \textsc{cmc} solution $\hat{\bm{\beta}}_{\alpha_n}=(\hat{{\beta}}_{\alpha_n}^0,\hat{{\beta}}_{\alpha_n}^1,\dots,\hat{{\beta}}_{\alpha_n}^p)^T$ as a classifier for the $p$ variables. We classify an $\mathbf{x}_j$ as active if $\hat{{\beta}}_{\alpha_n}^j\neq 0$ and inactive if $\hat{{\beta}}_{\alpha_n}^j=0$, so result (\ref{convg2}) also has the interpretation that the probability of no misclassification goes to 1. 

We refer to $\alpha_n$ as the $\alpha$-level of the \textsc{cmc}. It controls the balance between the false active rate and false inactive rate in the variable classification process. To see this, the \textsc{cmc} solution $\hat{\bm{\beta}}_{\alpha_n}$ is the minimizer of the model size under the constraint $\lambda(\hat{\bm{\beta}}_j) \leq \kappa_n$ and $\kappa_n$ depends on $\alpha_n$. By (\ref{kappa}), when $\alpha_n$ increases $\kappa_n$ decreases, so the minimum is taken over an increasingly smaller set of $\hat{\bm{\beta}}_j$ values and thus gets larger. The selected model will therefore have more and more variables. At the extreme value of $\alpha_n=1$, $\kappa_n=0$ and the \textsc{cmc} solution is the least squares estimator $\hat{\bm{\beta}}$ for the full model which, with probability one, has no zeros in its elements. Thus, the \textsc{cmc} selection is the full model with a 100\% false active rate and a 0\% false inactive rate. At the other extreme value of $\alpha_n=0$, $\kappa_n=\infty$ and the \textsc{cmc} selection is the empty model containing no variables, so it has a 0\% false active rate and a 100\% false inactive rate. By empirically exploring different $\alpha$-levels between 0 and 1, we may be able to find one that gives the desired balance between the two rates. In real applications where the sample size $n$ is fixed, we usually set the $\alpha$-level to a convenient value such as $0.1$, $0.5$ or $0.9$. See examples below.

We now present a numerical comparison on the accuracies of the \textsc{cmc}, the adjusted $R^2$, Mallows' $C_p$ which is equivalent to the \textsc{aic} in the present context, and the \textsc{bic}. We also provide recommendations on selecting the $\alpha$-level for the \textsc{cmc} based on simulation results.  The model used for this comparison is
\begin{equation}
\mathbf{y}=\beta_0 \mathbf{1}+ \mathbf{X}'\bm{\beta}' +\bm{\varepsilon}, \label{m01}
\end{equation}
where $\boldsymbol{\varepsilon} \sim N(\mathbf{0}, \sigma^2\mathbf{I})$ with $\sigma^2=1$, $\mathbf{X}'=[\mathbf{x}_1,\dots,\mathbf{x}_{p}]$, and the parameter values are $\beta_0=1$ and $\boldsymbol{\beta}'=(\beta_1,\dots,\beta_{p^*}, 0,\dots, 0)^T$ with $\beta_1=\dots=\beta_{p^*}=1$, so only the first $p^*$ variables are active.  We first consider the case where elements in $\mathbf{X}'$ are random numbers generated from a standard normal distribution. We call this the case of weakly correlated predictor variables as the observed correlations of the generated variables are small.
Table 1 contains values of the (false inactive rate, false active rate) pairs for six model selection criteria at 12 different combinations of $n$, $p$ and $p^*$.  The false inactive rate is the observed proportion of active variables misclassified as inactive and the false active rate is that of inactive variables misclassified as active, so when both rates are zero the selected model is the true model. Table 1 is computed using R package ``leaps'' by Lumley (2020) and each entry pair is based on 1000 simulation runs. For each run, we first generate an $\mathbf{X}'$ matrix and then through equation (\ref{m01}) a random $\mathbf{y}$ vector. We then perform the best subset selection using $\mathbf{X}'$ and  $\mathbf{y}$ with the six criteria to find their chosen models and compute their (false inactive rate, false active rate) pairs based on the chosen models. We repeat this process 1000 times and Table 1 records the average of the 1000 rate pairs. The subscript $\alpha$ in \textsc{cmc}$_\alpha$ is the $\alpha$-level for the \textsc{cmc}. We make the following observations based on results in Table 1:

\begin{table}
\caption{\label{tb-1} Variable classification accuracy comparison for the case of weakly correlated predictor variables: the entries are simulated (false inactive rate, false active rate) of the adjusted $R^2$, Mallows' $C_p$ (\textsc{aic}), \textsc{bic} and three \textsc{cmc}$_\alpha$ criteria for 12 scenarios. The bold  \textsc{cmc} results are those at recommended  $\alpha$-levels. Results in the table are the average of the two rates for 1000 simulation runs rounded to the second digit after the decimal point, and (0.00, 0.00)* means both rates are zero in all 1000 runs.}
\centering
{\small
\begin{tabular}{lcccccc} \\
$(n, \hspace{0.03in} p, \hspace{0.05in} p^*)$ & Adj $R^2$ & $C_p$ (\textsc{aic}) & \textsc{bic} &\textsc{cmc}$_{0.9}$ & \textsc{cmc}$_{0.5}$ & \textsc{cmc}$_{0.1}$ \\ \hline %\\[0.01pt]
%& \multicolumn{2}{c}{} & \multicolumn{2}{c}{} \\%[0.1in]
(20, 10, 5)   & (0.03, 0.39) & (0.06, 0.21) & (0.05, 0.24) & {\bf (0.07, 0.16)}  & (0.18, 0.07) & (0.41, 0.03) \\
(30, 10, 5)   & (0.00, 0.35) & (0.00, 0.18) & (0.01, 0.13) & (0.01, 0.11) & {\bf (0.04, 0.03)} & (0.17, 0.01)\\
(40, 10, 5)   & (0.00, 0.32) & (0.00, 0.17) & (0.00, 0.09) & (0.00, 0.10) & {\bf (0.01, 0.02)} & (0.06, 0.00)\\
(50, 10, 5)   & (0.00, 0.32) & (0.00, 0.18) & (0.00, 0.08) & (0.00, 0.09) & (0.00, 0.02) & {\bf (0.01, 0.00)}\\  \hline %\\[0.01pt]
% an extreme case showing the probability that the LS estimator for the true model is in the CR does not go to 1.
%(100000, 10, 5)   & (0, 0.001) & (0, 0.166) & (0, 0.001) & (0, 0.085) & (0, 0.017) & (0, 0.001)

(40, 20, 10)   & (0.00, 0.35) & (0.00, 0.18) & (0.00, 0.15) & {\bf (0.01, 0.06)} & (0.06, 0.02) & (0.17, 0.02)\\
(60, 20, 10)   & (0.00, 0.30) & (0.00, 0.17) & (0.00, 0.08) & (0.00, 0.04) & {\bf (0.01, 0.01)} & (0.04, 0.00)\\
(80, 20, 10)   & (0.00, 0.28) & (0.00, 0.16) & (0.00, 0.06) & (0.00, 0.03) & {\bf (0.00, 0.00)} & (0.00, 0.00)\\
(100, 20, 10)   & (0.00, 0.26) & (0.00, 0.16) & (0.00, 0.05) & (0.00, 0.03) & (0.00, 0.00) & {\bf (0.00, 0.00)*} \\  \hline %\\[0.01pt]

(60, 30, 15)   & (0.00, 0.31) & (0.00, 0.17) & (0.00, 0.11) & {\bf (0.00, 0.03)} & (0.02, 0.01) & (0.08, 0.00)\\
(90, 30, 15)   & (0.00, 0.27) & (0.00, 0.16) & (0.00, 0.07) & (0.00, 0.02) & {\bf (0.00, 0.01)} & (0.01, 0.00)\\
(120, 30, 15)   & (0.00, 0.24) & (0.00, 0.16) & (0.00, 0.04) & (0.00, 0.01) & {\bf (0.00, 0.00)} & (0.00, 0.00)\\
(150, 30, 15)   & (0.00, 0.22) & (0.00, 0.15) & (0.00, 0.04) & (0.00, 0.01) & (0.00, 0.00) & {\bf (0.00, 0.00)*}

\end{tabular}
}
\end{table}

\begin{itemize}

\item[1.] When the sample size $n$ is not large relative to the number of variables $p$  (the three cases in Table 1 where $n=2p$), none of the six criteria performed very well. The adjusted $R^2$, $C_p$ and \textsc{bic} have low false inactive rates but high false active rates. The \textsc{cmc}$_{0.5}$ and  \textsc{cmc}$_{0.1}$  have high false inactive rates but low false active rates. If we treat false active and false inactive as equally serious errors and rank the six criteria by the overall error rate (defined as the sum of the two error rates), then the \textsc{cmc}$_{0.9}$ is the best of the three \textsc{cmc} criteria.  For such small sample cases, the $\alpha$-level of the \textsc{cmc} needs to be set to a high level or the false inactive rate and the overall error rate could be high.
\item[2.] When $n$ is large relative to $p$  (cases where $n\geq 3p$), the \textsc{cmc} is very accurate. While all 6 criteria have zero or low false inactive rates, the three \textsc{cmc} criteria also have low false active rates, whereas the adjusted $R^2$ and $C_p$ still have double-digit false active rates. For such large sample cases, the \textsc{cmc} outperforms the other three criteria in terms of the overall error rate. In particular, overall the \textsc{cmc}$_{0.5}$ is the most accurate among the six criteria.
\item[3.] Focusing now on the $\alpha$-level of the \textsc{cmc}, we have noted that the false inactive rate decreases and the false active rate increases when the $\alpha$-level increases. We see from Table 1 that for any given $(n,p,p^*)$ combination, this is indeed the case. Based on our experience for the case of weakly correlated predictor variables, to achieve a small overall error rate we recommend setting the $\alpha$-level to 0.9 if $n\leq 2p$, to 0.5 if $2p<n\leq 4p$, and to 0.1 only when $n > 4p$.  The  \textsc{cmc} results for such recommended levels are in bold font in Table 1. 
In real data analysis, it may be helpful to use two levels at the same time. See Section 3 for a real example and Section 4 for a discussion about this point.  

\end{itemize}

Table \ref{tb-2} constructed using relevant results from Table \ref{tb-1} illustrates the consistency of the \textsc{cmc} selection in (\ref{convg2}) that the probability of no classification errors approaches one.
It shows that as $n$ increases from 40 to 100 and $\alpha_n$ decreases from 0.90 to 0.10, the false inactive and false active rates both go down towards zero as predicted by  (\ref{convg2}). Noting that the rate values in Tables \ref{tb-1} and \ref{tb-2} are the averages of such rates in 1000 runs, the pair of zero rates for the case of $n=100$ and $\alpha_n=0.10$ implies none of the underlying 1000 runs had any classification error, i.e., the \textsc{cmc} selection is 100\% accurate in these runs. In practice, $\alpha_n$ does not have to be very small in order for the \textsc{cmc} to be very accurate so long as the sample size is large; e.g., for the cases of $\alpha_n= 0.5$ in Table \ref{tb-1}, we also observed zero or near zero error rates at large sample sizes of $n\geq 3p$. It is worth noting that the \textsc{cmc} is the only criterion that reached zero error rates at $n=5p$ for $p=20, 30$. While Adjusted $R^2$ and $C_p$ criteria are not expected to reach zero rates, the \textsc{bic} is consistent in the current setting and it should reach zero rates as $n$ goes to infinity. The observation that it did not even at $n=5p$ suggests that the error rates of the \textsc{bic} converge to zero at a slower rate than that of the \textsc{cmc}.

\begin{table}
\caption{\label{tb-2}  An example illustrating the consistency of the \textsc{cmc} (\ref{convg2}): the probability of false inactive rate (FIR) and false active rate (FAR) both being zero approaches 1 as $n\rightarrow \infty$ and $\alpha_n \rightarrow 0$.}
\centering
{\small
\begin{tabular}{ccc} \\
$(n, \hspace{0.03in} p, \hspace{0.05in} p^*)$ & $\alpha_n$ & (FIR, FAR) \\ \hline 
%(20, 10, 5)   & 0.50 & (0.19, 0.07) \\
%(40, 10, 5)   & 0.80 & (0.03, 0.01) \\
%(60, 10, 5)   & 0.90 & (0.01, 0.00)\\  \hline %\\[0.01pt]

(40, 20, 10)   & 0.90 & (0.01, 0.06) \\
(60, 20, 10)   & 0.50 & (0.01, 0.01) \\
(100, 20, 10)   & 0.10 & (0.00, 0.00)*
\end{tabular}
}
\end{table}

We now consider a case where there are systematic correlations among the predictor variables and examine how such correlations affect the performance of the model selection criteria. For this case, we set $(p,p^*)=(20,10)$ so that in model (\ref{m01}) $\mathbf{x}_1, \dots, \mathbf{x}_{10}$ are active variables and  $\mathbf{x}_{11}, \dots, \mathbf{x}_{20}$ are inactive variables. We then introduce correlations among 5 active variables and do the same for 5 inactive variables as follows. Let $\mathbf{z}_1, \dots, \mathbf{z}_{22}$ be 22 independent $n$-variate standard normal random vectors. For some $w\in (0,1)$, define

 \begin{centering}
 $\mathbf{x}_i=(1-w)\mathbf{z}_i+w\mathbf{z}_{21}$ for $i=1, \dots, 5$ and $\mathbf{x}_i=\mathbf{z}_{i}$ for $i=6, \dots, 10$; and\\
 $\mathbf{x}_i=(1-w)\mathbf{z}_i+w\mathbf{z}_{22}$ for $i=11, \dots, 15$, and $\mathbf{x}_i=\mathbf{z}_{i}$ for $i=16, \dots, 20$,\\
 \end{centering}

\noindent so the first 5 active variables are correlated with a theoretical pairwise correlation coefficient of $\rho=w^2/[(1-w)^2+w^2]$. The first 5 inactive variables also have the same pairwise correlation coefficient. When we generate the $\mathbf{x}_i$ through the $\mathbf{z}_i$, the observed correlation coefficients between the correlated $\mathbf{x}_i$ are approximately $\rho$ which goes to 1 as $w$ goes to 1. This allows us to simulate predictor variables with different degrees of correlation by changing the $w$ value. Table 3 contains simulated (false inactive rate, false active rate) of the adjusted $R^2$, Mallows' $C_p$ (\textsc{aic}), \textsc{bic} and three \textsc{cmc}$_\alpha$ criteria for 3 levels of correlations $\rho=0.3, 0.5, 0.8$, representing moderate, medium and strong levels of correlation. With half of the variables correlated, the accuracies of all criteria are negatively affected when compared to the weakly correlated case in the middle section of Table 1 where $(p, p^*)$ is also $(20,10)$. For all six criteria, the biggest impact of the correlation in predictor variables is an increase in false inactive rates, especially at $\rho=0.8$, when compared to the case of weak correlation in Table 1. This is not unexpected as it is known that when active variables are strongly correlated, some of them tend to be dropped in the  variable selection process. For the \textsc{cmc}, this increase in the false inactive rate slows the speed at the which the two error rates converge to zero; e.g., in the middle section of Table 1, the rates reached zero when $n=5p$, but in Table 3 when $\rho=0.8$ the rates reached zero only when $n=20p$. Other examples containing a large number of moderately to strongly correlated variables that we have studied also showed a similar slowing in convergence to zero. Further, for $n\leq 5p$ the most accurate \textsc{cmc} level is often the 0.9 level. Based on these findings, for situations with moderately to strongly correlated variables we recommend setting the $\alpha$-level to 0.9 if $n\leq 5p$, and to 0.5 if $5p<n \leq 15p$. Level 0.1 may be used when  $n> 15p$.

\begin{table}
\caption{\label{tb-3} Variable classification accuracy comparison for a case of correlated predictor variables: the entries are simulated (false inactive rate, false active rate) of the adjusted $R^2$, Mallows' $C_p$ (\textsc{aic}), \textsc{bic} and three \textsc{cmc}$_\alpha$ criteria. The bold \textsc{cmc} results are those at recommended  \textsc{cmc} levels. Results in the table are the average of the two rates for 1000 simulation runs for the case of $(p, p^*)=(20, 10)$.}
\centering
{\small
\begin{tabular}{lcccccc} \\
$(\rho, \hspace{0.03in} n)$ & Adj $R^2$ & $C_p$ (\textsc{aic}) & \textsc{bic} &\textsc{cmc}$_{0.9}$ & \textsc{cmc}$_{0.5}$ & \textsc{cmc}$_{0.1}$ \\ \hline %\\[0.01pt]
%& \multicolumn{2}{c}{} & \multicolumn{2}{c}{} \\%[0.1in]
(0.3, 40)   & (0.02, 0.35) & (0.03, 0.18) & (0.03, 0.15) & {\bf (0.06, 0.07)}  & (0.14, 0.03) & (0.26, 0.01) \\
(0.3, 60)   & (0.00, 0.30) & (0.00, 0.17) & (0.01, 0.09) & {\bf (0.01, 0.04)} & {(0.06, 0.01)} & (0.13, 0.00)\\
(0.3, 100)  & (0.00, 0.26) & (0.00, 0.16) & (0.00, 0.05) & {\bf (0.00, 0.03)} & {(0.01, 0.00)} & {(0.04, 0.00)}\\ 
(0.3, 200)  & (0.00, 0.20) & (0.00, 0.15) & (0.00, 0.03) & { (0.00, 0.03)} & {\bf {(0.00, 0.00)}} & {(0.00, 0.00)}\\  \hline %\\[0.01pt]

(0.5, 40)   & (0.04, 0.35) & (0.06, 0.18) & (0.07, 0.16) & {\bf (0.11, 0.08)}  & (0.18, 0.03) & (0.30, 0.02) \\
(0.5, 60)   & (0.00, 0.29) & (0.00, 0.17) & (0.02, 0.09) & {\bf (0.04, 0.05)} & {(0.90, 0.02)} & (0.17, 0.00)\\
(0.5, 100)  & (0.00, 0.26) & (0.00, 0.17) & (0.00, 0.05) & {\bf (0.00, 0.03)} & {(0.02, 0.01)} & {(0.08, 0.00)}\\
(0.5, 200)  & (0.00, 0.20) & (0.00, 0.16) & (0.00, 0.03) & { (0.00, 0.03)} & {\bf (0.00, 0.00)} & {(0.00, 0.00)}\\  \hline %\\[0.01pt]

(0.8, 40)   & (0.11, 0.36) & (0.15, 0.20) & (0.16, 0.16) & {\bf (0.19, 0.09)}  & (0.26, 0.03) & (0.36, 0.01) \\
(0.8, 60)   & (0.04, 0.31) & (0.07, 0.18) & (0.10, 0.09) & {\bf (0.11, 0.07)} & {(0.18, 0.02)} & (0.25, 0.00)\\
(0.8, 100)  & (0.01, 0.24) & (0.01, 0.16) & (0.04, 0.05) & {\bf (0.05, 0.04)} & (0.10, 0.01) & {(0.16, 0.00)}\\ 
(0.8, 200)  & (0.00, 0.18) & (0.00, 0.16) & (0.00, 0.03) & {(0.00, 0.03)} & {\bf (0.02, 0.00)} & {(0.08, 0.00)}\\ 
(0.8, 400)  & (0.00, 0.12) & (0.00, 0.16) & (0.00, 0.02) & (0.00, 0.03) & {(0.00, 0.00)} & {\bf (0.00, 0.00)}\\  \hline %\\[0.01pt]

\end{tabular}
}
\end{table}

In practice, whether or not there are correlated predictor variables may be determined by examining the correlation matrix of the variables. If one is unsure after examining the correlation matrix or wishes to avoid looking at this issue when applying the \textsc{cmc}, we recommend the 0.9 level as the default $\alpha$-level for such situations. Numerical results such as those in Tables 1 and 2 show that at this level, the \textsc{cmc} is either similar to or better than the \textsc{bic} in terms of the overall error rate regardless the underlying correlations of the variables and sample size, and it is consistently better than the adjusted $R^2$ and Mallows' $C_p$ (\textsc{aic}). The availability of such a default level makes the \textsc{cmc} a practical alternative to other criteria as a user does not have to fine-tune its $\alpha$-level in order to make \textsc{cmc} competitive; the \textsc{cmc} under the default level already has better performance than other criteria. Nevertheless, if we want to realize the full potential of the \textsc{cmc}, we need to optimize the choice of its $\alpha$-level by examining the correlations of the predictor variables and following the detailed recommendations about the selection of the $\alpha$-level given above.

Finally, for brevity of presentation we have only presented examples where parameters of active variables are all 1 and $p^*=p/2$. We have tried different $p^*$ and used parameter values that differ in size and sign, and obtained similar observations concerning the relative performance of the six criteria.

%For situations with a large number of strongly correlated variables, use only level 0.9 when $n\leq 5p$ and smaller levels may be used when $n>5p$.  If there is a silver lining in the speed of the \textsc{cmc} here, it is that at level 0.1 it consistently reaches zero rates faster than the only other consistent criterion, the \textsc{bic}, regardless the correlation situation; e.g., for the example containing a large number of strongly correlated variables of $\rho=0.8$ in Table 3, at $n=400$ and $\alpha=0.1$ the two rates of the \textsc{cmc} are zero but that of the false active rate of the \textsc{bic} is still about 2%.

%33333333333333333333333333333333333333333333333333333333333333333333333333333333333333333333333333333333333333333333333333333333333333333

\section{Application to a prostate cancer data}

We now apply the \textsc{cmc} to analyse a prostate cancer data set from Stamey et al. (1989). This data set had been analysed in the book by Hastie, Tibshirani  and Friedman (2009). It is concerned with the correlation between the level of prostate specific antigen (PSA) and a number of clinical measures in 97 men who were about to receive a radical prostatectomy. There are 9 variables in the data set: the log of PSA (lpsa), log cancer volume (lcavol), log prostate weight (lweight), age, log of benign prostatic hyperplasia amount (lbph), seminal vesicle invasion (svi), log of capsular penetration (lcp), Gleason score (gleason), and percent of Gleason scores 4 or 5 (pgg45). The goal is to predict the lpsa by using the other 8 variables through a linear model. Scatter plots of the variables can be found in Figure 1.1 of the book which suggest a linear model is appropriate. Table 3.3 of the book contains the estimated model based on measurements from a subset of 67 men by six different methods: the ordinary least squares regression, the best subset selection, ridge regression, the lasso, principle component regression and partial least squares regression. The model selection and penalty parameter selection criterion used for the last five methods is cross-validation. To compute Table 3.3, the 8 predictor variables were all standardized.

\begin{table}
\caption{\label{tb-4} Estimated models by the best subset selection method with five variable selection criteria. The \textsc{cmc}$_{0.1}$, \textsc{cmc}$_{0.5}$ and \textsc{bic} selected the same model. Notation: ``***": significant at 0.001 level; ``**": significant at 0.01 level; ``*": significant at 0.05 level.}
\centering
\begin{tabular}{llccccc} \\

Variable  & LS    & Adj $R^2$ & $C_p$  & \textsc{bic}    &\textsc{cmc}$_{0.5}$  & \textsc{cmc}$_{0.1}$ \\ \hline
Intercept & 2.478 *** & 2.478     & 2.478  & 2.478  & 2.478      &  2.478  \\
lcavol    & 0.665 *** & 0.671     & 0.622  & 0.619  & 0.619      & 0.619  \\
lweight   & 0.266 ** & 0.263     & 0.229  & 0.283  & 0.283      & 0.283  \\
age       & $-0.158$   &$ -0.155$    &        &        &            &        \\
lbph      & 0.140    & 0.141     & 0.114  &        &            &           \\
svi       & 0.315 ** & 0.311     & 0.292  & 0.275  & 0.275      &  0.275    \\
lcp       & $-0.148$   & $-0.146$    &        &        &            &            \\
gleason   & 0.035    &           &        &        &            &          \\
pgg45     & 0.125    & 0.150     &        &        &            &         \\
\end{tabular}
\end{table}

Since we do not need to split the sample for cross-validation, we use all 97 observations in the best subset selection. Due to the large sample size relative to the number of variables and no strong correlations were found among the 8 variables, we choose to use $\alpha$-levels 0.1 and 0.5. To facilitate comparison with Table 3.3 in Hastie, Tibshirani  and Friedman (2009), we also standardized the predictor variables. Table \ref{tb-4} contains models chosen by five criteria. It shows that  \textsc{cmc}$_{0.1}$, \textsc{cmc}$_{0.5}$ and the \textsc{bic} selected the same model with 3 variables lcavol+lweight+svi. The $C_p$ criterion selected one more variable, lbph. The adjusted $R^2$ criterion dropped only one variable, gleason, from the full model.
Comparing Table \ref{tb-4} with Table 3.3 in Hastie, Tibshirani  and Friedman (2009), we see that there are some differences in the least squares estimates which are due to our use of the full data set for estimation and model selection, but estimates for parameters that are significant are approximately equal. While the best subset selection based on the cross-validation in Table 3.3 selected only lcavol+lweight, that based on the five criteria in Table \ref{tb-4} all selected more variables. In particular, the three-variable model chosen by \textsc{cmc}$_{0.1}$, \textsc{cmc}$_{0.5}$ and \textsc{bic} has one more variable than the best subset selection in Table 3.3 and one fewer than the lasso selection in Table 3.3 which is lcavol+lweight+svi+lbph. Since lbph is not significant in the full model, the estimated parameter value for lbph in the lasso selection is only 0.002 (which is substantially smaller than other values) and it is known that the lasso with cross-validation tends to include false active variables, lbph is likely a false active variable. So our final selection is the three-variable model  lcavol+lweight+svi chosen by the two \textsc{cmc} criteria and the \textsc{bic}.

\section{Concluding remarks}

The \textsc{cmc}'s ability to control the balance between the false active rate and false inactive rate through its $\alpha$-level is rooted in the underlying hypothesis test's ability to control the Type-I and Type-II errors through its significance level. This feature of the \textsc{cmc} is particularly useful when one type of misclassification is considered more serious than another and we want to keep the rate of the more serious misclassification to a low level which can be achieved by adjusting the $\alpha$-level. For real applications, we may consider one or two $\alpha$-levels besides the recommended level and compare their \textsc{cmc} selections to make a final choice. Since a higher $\alpha$-level gives the \textsc{cmc} selection a lower false inactive rate and a lower level gives it a lower false active rate, if and when both levels lead to the same selection as in the prostate cancer data example, we have added confidence in the accuracy of the \textsc{cmc} selection as both rates associated with this selection are expected to be low. This ability is an advantage to the \textsc{cmc} as other model selection criteria do not have control on the balance between the two rates.

The best subset selection used to be only viable for model selection problems with $p<40$ due to its high computational cost. Recently, however, Bertsimas, King and Mazumder (2016) developed an efficient way to perform the best subset selection for $p$ in the hundreds and even thousands. Thus, the \textsc{cmc} may be applied to perform the best subset selection for problems with a large $p$. When the $p$ is too large and the best subset selection becomes impractical, we may apply the \textsc{cmc} with the lasso or variants of the lasso discussed in Hastie, Tibshirani and Wainwright (2015). When applied with the lasso, for example, the set of models under consideration is the sequence of models found along the lasso path which we denote with  ${\cal M}_L$, so the minimization in (\ref{form1}) is done over ${\cal M}_L$, not over the set of all possible models.  The proof of Theorem 2.1 can be modified to show that in this case the \textsc{cmc} solution is still a consistent estimator for $ \bm{\beta}^t $ and, under an extra condition that the true model is in ${\cal M}_L$ with probability tending to one as the sample size increases, the \textsc{cmc} selection is also consistent.

The \textsc{cmc} requires the condition that $n>p$ to ensure the least squares estimators $\hat{\bm{\beta}}_j$ and their likelihood ratios are all well-defined so that the likelihood ratio test is valid. For high dimensional situations where $p>n$, we may reduce the dimension using the lasso before applying the \textsc{cmc}. To do so, we first use the lasso with the cross-validation criterion to select a model and then use this model as the ``full model'' for the best subset selection with the  \textsc{cmc}.  This ``full model'' has fewer than $n$ variables and it tends to have a high number of false active variables. Simulation results (not included here) show that the \textsc{cmc} is very effective at reducing its number of false active variables.

To conclude, the \textsc{cmc} may be extended to handle model selection problems for other types of models such as generalized linear models. To define a \textsc{cmc} for a model selection problem, the key ingredient we need is a powerful test for individual models with a corresponding Hodges-Lehmann estimator that is consistent and a corresponding confidence region that can be kept at $o_p(n)$ size when the confidence level approaches one.  When there is such a powerful test, the resulting \textsc{cmc} would be a strong competitor to the existing criteria.

%\begin{appendix}

\section*{Appendix}
\label{App}

Under model (\ref{m0}), the log-likelihood function for $\bm{\beta}$ is
\beq
l({\boldsymbol{\beta}})=-\frac{n}{2}\ln(2\pi) -n\ln (\sigma)-
\frac{1}{2\sigma^2}\|\mathbf{y}-\mathbf{X}\bm{\beta}\|_2^2,  \label{logl}
\eeq
which is maximized at $\bm{\beta}=\hat{\bm{\beta}}$ for any fixed $\sigma^2$. The likelihood ratio statistic for testing the null hypothesis $H_0:$ ${\bm{\beta}^t=\bm{\beta}}$ is
\beq
-2\{l({\boldsymbol{\beta}})-l({\hat{\bm{\beta}}})\}=
\frac{1}{{\sigma}^2}\left\{ \|\mathbf{y}-\mathbf{X}\bm{\beta}\|_2^2-\|\mathbf{y}-\mathbf{X}\hat{\bm{\beta}}\|_2^2\right\} \label{lrs_1}
\eeq
which contains the unknown parameter $\sigma^2$. For model selection, $\sigma^2$ is a nuisance parameter and it does not affect the relative likelihood of different $\bm{\beta}$ values. Let $\hat{\sigma}^2$ be the residual mean square from the least squares regression for the full model (\ref{m0}). Since $\hat{\sigma}^2$ is a consistent estimator for $\sigma^2$, we substitute it for the unknown $\sigma^2$ in (\ref{lrs_1}) and the likelihood ratio test statistic becomes
\beq
\lambda({\boldsymbol{\beta}})= 
\frac{1}{\hat{\sigma}^2}\left\{ \|\mathbf{y}-\mathbf{X}\bm{\beta}\|_2^2-\|\mathbf{y}-\mathbf{X}\hat{\bm{\beta}}\|_2^2\right\}.  \label{lrs}
\eeq
Under $H_0$, the asymptotic distribution of a likelihood ratio test statistic involving an estimated parameter is often a scaled or a weighted $\chi^2$ distribution. That of $\lambda({\boldsymbol{\beta}})$ in (\ref{lrs}) is also a scaled $\chi^2$ distribution, but since the exact finite sample distribution of $\lambda({\boldsymbol{\beta}})$ is available, we will use the exact distribution.

\vspace{0.2in}

\noindent {\bf Proof of Theorem 2.1.} 
It may be verified that the likelihood ratio statistic $\lambda({\bm{\beta}})$ defined in (\ref{lrs}) can be expressed as
\beq
\lambda({\bm{\beta}})=
\frac{1}{\hat{\sigma}^2}\left\{ ({\bm{\beta}}-\hat{\boldsymbol{\beta}})^T\mathbf{X}^T\mathbf{X}({\bm{\beta}}-\hat{\boldsymbol{\beta}})\right\}, \label{lr}
\eeq
and that under $H_0:$ ${\bm{\beta}^t=\bm{\beta}}$ the distribution of $q^{-1}\lambda({\bm{\beta}})$ is an $F_{q, n-q}$ distribution. Thus, the collection of ${\bm{\beta}}$ values not rejected by the likelihood ratio test at the $\alpha_n$ level form the following $100(1-\alpha_n)\%$ confidence region for $\bm{\beta}^t$,
\beq
{\cal C}_{\alpha_n} = \left\{ \bm{\beta}: \frac{ (\bm{\beta}-\hat{\boldsymbol{\beta}})^T\mathbf{X}^T\mathbf{X}(\bm{\beta}-\hat{\boldsymbol{\beta}}) }{  q\hat{\sigma}^2} \leq F_{{1-\alpha_n},q, n-q}\right\}.  \label{cr}
\eeq
By (\ref{kappa}) and (\ref{lr}), we may express ${\cal C}_{\alpha_n}$  in terms of $\lambda({\boldsymbol{\beta}})$ and $\kappa_n$ as
\beq
{\cal C}_{\alpha_n}=\{ \bm{\beta}: \lambda({\boldsymbol{\beta}})\leq \kappa_n\}. \label{equiv}
\eeq
Since the \textsc{cmc} solution $\hat{\bm{\beta}}_{\alpha_n}$ for (\ref{form1}) satisfies $\lambda(\hat{\bm{\beta}}_{\alpha_n})\leq \kappa_n$, (\ref{equiv}) implies that $\hat{\bm{\beta}}_{\alpha_n}\in {\cal C}_{\alpha_n}$. This, (\ref{cr}) and $F_{1-\alpha_n,q, n-q}=o(n)$ imply
\beq
\frac{1}{n}(\hat{\bm{\beta}}_{\alpha_n}-\hat{\boldsymbol{\beta}})^T\mathbf{X}^T\mathbf{X}(\hat{\bm{\beta}}_{\alpha_n}-\hat{\boldsymbol{\beta}})
  \leq \frac{1}{n}q\hat{\sigma}^2 F_{1-\alpha_n,q, n-q} = o_p(1),  \label{ineq1}
\eeq
where the right-hand side is $o_p(1)$ instead of $o(1)$ because $\hat{\sigma}^2$ is not a constant but it converges in probability to $\sigma^2$. 
By (\ref{cond1}) and (\ref{ineq1}),  $\|\hat{\bm{\beta}}_{\alpha_n}-\hat{\bm{\beta}}\|_2^2=o_p(1)$. Also, by
(\ref{asynorm}) $\|\hat{\bm{\beta}}-\bm{\beta}^t\|_2^2=O_p(n^{-1})$. It follows that
\beq
\|\hat{\bm{\beta}}_{\alpha_n}-\bm{\beta}^t\|_2^2\leq 
\|\hat{\bm{\beta}}_{\alpha_n}-\hat{\bm{\beta}}\|_2^2 + \|\hat{\bm{\beta}}-\bm{\beta}^t\|_2^2 =o_p(1),
\label{op1}
\eeq
which proves the consistency of the \textsc{cmc} solution (\ref{convg1}).

To prove the \textsc{cmc} selection consistency (\ref{convg2}), we first examine which elements of vectors $\bm{\beta}$ in ${\cal C}_{\alpha_n}$ may be zero and which elements may not be zero when $n$ is large. Define an event
\[ \mbox{
$A=$ \{Elements of $\bm{\beta}$ in ${\cal C}_{\alpha_n}$ corresponding to non-zero elements of $\bm{\beta}^t$ are also non-zero\}. }
\]
By the definition of ${\cal C}_{\alpha_n}$ in (\ref{cr}), vectors $\bm{\beta}$ in ${\cal C}_{\alpha_n}$ satisfy
\beq
\frac{1}{n}({\bm{\beta}}-\hat{\boldsymbol{\beta}})^T\mathbf{X}^T\mathbf{X}({\bm{\beta}}-\hat{\boldsymbol{\beta}})
  \leq \frac{1}{n}q\hat{\sigma}^2 F_{1-\alpha_n,q, n-q} = o_p(1),  \label{ineq2}
\eeq
uniformly. Similar to (\ref{op1}), by the triangular inequality we have $\|{\bm{\beta}}-\bm{\beta}^t\|_2^2 =o_p(1)$ uniformly for all  $\bm{\beta} \in {\cal C}_{\alpha_n}$. This implies individual elements of $\bm{\beta}$ converge in probability to corresponding elements of $\bm{\beta}^t$ uniformly and thus $P(A) \rightarrow 1$  as the sample size $n$ goes to infinity. When event $A$ occurs, among the set of vectors $\{{\hat{\bm{\beta}}_j} \}_{j=1}^{2^p}$ only those for models containing all active variables can be in ${\cal C}_{\alpha_n}$, so $\hat{\bm{\beta}}_{j}^*$ of the true model $M^*_j$ would be the most sparse member of $\{{\hat{\bm{\beta}}_j} \}_{j=1}^{2^p}$ that may possibly be in ${\cal C}_{\alpha_n}$. It follows that $\{\hat{\bm{\beta}}_{j}^* \in {\cal C}_{\alpha_n}\}\cap {A}$ implies $\{\hat{\bm{\beta}}_{\alpha_n} = \hat{\bm{\beta}}_{j}^*\}$, so
\beq
P(\hat{\bm{\beta}}_{\alpha_n} = \hat{\bm{\beta}}_{j}^*) \geq
P( \{\hat{\bm{\beta}}_{j}^* \in {\cal C}_{\alpha_n}\} \cap {A}) \rightarrow P( \hat{\bm{\beta}}_{j}^* \in {\cal C}_{\alpha_n})  \label{temp1}
\eeq
as $n$ goes to infinity. Also, $\hat{\bm{\beta}}_{j}^* \in {\cal C}_{\alpha_n}$ if ${\bm{\beta}^t}\in {\cal C}_{\alpha_n}$ since $\hat{\bm{\beta}}_{j}^*$ has a higher likelihood and thus a smaller likelihood ratio than ${\bm{\beta}^t}$. It follows that
\beq
P( \hat{\bm{\beta}}_{j}^* \in {\cal C}_{\alpha_n})\geq P(\bm{\beta}^t \in {\cal C}_{\alpha_n}) = 1-\alpha_n \rightarrow 1.  \label{temp2}
\eeq
Equations (\ref{temp1}) and (\ref{temp2}) then imply (\ref{convg2}).  \hfill $\Box$

\vspace{0.2in}

We now remark on the condition of $\alpha_n \rightarrow 0$ and $F_{1-\alpha_n,q, n-q}=o(n)$ in Theorem 2.1. It requires that $\alpha_n$ approaches 0 slowly so that $F_{1-\alpha_n,q, n-q}/n$ approaches 0 when $n$ goes to infinity. A sequence of $\alpha_n$ satisfying this condition can be constructed in many ways, e.g., by letting 
\beq 
1-\alpha_n=P(F_{q,n-q}\leq n^{\delta}) \label{gseq}
\eeq
where $\delta$ is a small positive number so that $F_{1-\alpha_n,q, n-q}=n^{\delta}=o(n)$. Since $F_{q,n-q}$ converges to $\chi^2_q/q$ as $n$ goes to infinity,  (\ref{gseq}) implies that when $n$ is large $\alpha_n$ is essentially $1-P(\chi^2_{q}\leq qn^{\delta})$ which satisfies $\alpha_n\rightarrow 0$. In real applications where $n$ is fixed, we do not choose the $\alpha_n$ using  (\ref{gseq}). Instead, we set it to convenient values such as 0.1, 0.5 and 0.9 depending on the relative size of $n$ and $p$ and whether there are correlated variables, or we simply set it to the default value of 0.9, as we recommended in Section 2. Nevertheless, the condition is an important reminder that we should not let $\alpha_n$ to be too close to 0 when $n$ is not large. Otherwise, the \textsc{cmc} selection may be inaccurate. For example, the \textsc{cmc}$_{0.1}$ for the small sample case of $(20, 10, 5)$  in Table 1 has a high false inactive rate of 41\%. On the other hand, when the sample size is really large, very small $\alpha$-levels could be used. As an example of this, for all three combinations of $p$ and $p^*$ in Table 1, when $n=300$ the simulated \textsc{cmc} error rates at the 0.1, 0.05 and 0.01 levels are all zero, whereas the false active rate of the  \textsc{bic} is still around 2\% and that of the adjusted $R^2$ and $C_p$ are still in double digits.

We have observed in Section 2 that the $\alpha$-level does not have to be very small to obtain accurate results as long as the sample size is large. As a part of future research on the $\textsc{cmc}$, we will look for a theoretical explanation for this phenomenon which may lead to a more effective way of choosing the $\alpha$-level.
%\end{appendix}

%%%%%%%%%%%%%%%%%%%%%%%%%%%%%%%%%%%%%%%%%%%%%%%%%%%%%%%%%%%%%%%%%%%%%%%%%%%%%%%%%%%%%%%%%%%%%%%%%%%%%%%%%%%%%%%%%%%%%%%%%%%%
\vskip 14pt
%\noindent {\large\bf Supplementary Materials}

%The Supplementary Material is a self-contained document containing additional computational details and numerical examples, and %applications and discussions concerning the variability weighted average effect.
%\par
%%%%%%%%%%%%%%%%%%%%%%%%%%%%%%%%%%%%%%%%%%%%%%%%%%%%%%%%%%%%%%%%%%%%%%%%%%%%%%%%%%%%%%%%%%%%%%%%%%%%%%%%%%%%%%%%%%%%%%%%%%%%
%\vskip 14pt
%\noindent {\large\bf Acknowledgements}

%This work is supported by a research grant from the Southern University of Science and Technology of China and a Discovery Grant from the National Science and Engineering Research Council of Canada. We would like to thank two anonymous reviewers for comments that have led to improvements in this paper.

%\begin{ack}{ACKNOWLEDGEMENTS}

%\end{ack}

\end{document}